\documentclass[fleqn,10pt]{wlscirep}
\usepackage[utf8]{inputenc}
\usepackage[T1]{fontenc}

\usepackage{graphicx}
\usepackage{comment}
\newcommand{\heading}[1]{\noindent\textbf{#1}}

\usepackage{amsmath}
\usepackage{textcomp} 
\usepackage{amsfonts}       
\usepackage{url}

\usepackage{graphicx}
\usepackage{siunitx}
\usepackage{xcolor}
\usepackage{comment}
\usepackage{graphicx}
\usepackage{amsmath}
\usepackage{pifont}
\usepackage{amsmath, amssymb, amsfonts, amsthm}
\usepackage{graphicx}
\usepackage{amsfonts}       
\usepackage{nicefrac}       
\usepackage{microtype}      
\usepackage{mathtools}
\usepackage{comment}
\usepackage{pbox}
\usepackage{textcomp}
\usepackage[utf8]{inputenc}
\usepackage{multirow}
\usepackage{nameref}
\usepackage{textalpha}
\usepackage{booktabs} 
\usepackage{comment}
\usepackage{subfig}
\usepackage{threeparttable}
\usepackage{caption}

\usepackage{ragged2e} 

\usepackage{appendix}

\usepackage[normalem]{ulem} 
\newcommand{\stkout}[1]{\ifmmode\text{\sout{\ensuremath{#1}}}\else\sout{#1}\fi}

\newcommand{\deletedfloat}[1]{}

\title{Abnormality Prediction and Forecasting of Laboratory Values from Electrocardiogram Signals Using Multimodal Deep Learning}

\author[1]{Juan Miguel Lopez Alcaraz}
\author[1*]{Nils Strodthoff}
\affil[1]{Carl von Ossietzky Universität Oldenburg, AI4Health Division, Oldenburg, 26129, Germany}

\affil[*]{nils.strodthoff@uol.de}

\keywords{Electrocardiogram (ECG), Laboratory abnormalities, Multimodal deep learning, Structured state space model, Clinical decision support}

\begin{abstract}
This study investigates the feasibility of using electrocardiogram (ECG) data combined with basic patient metadata to estimate and monitor prompt laboratory abnormalities. We use the MIMIC-IV dataset to train multimodal deep learning models on ECG waveforms, demographics, biometrics, and vital signs. Our model is a structured state space classifier with late fusion for metadata. We frame the task as individual binary classifications per abnormality and evaluate performance using AUROC. The models achieve strong performance, with AUROCs above 0.70 for 24 lab values in abnormality prediction and up to 24 in abnormality forecasting, across cardiac, renal, hematological, metabolic, immunological, and coagulation categories. NTproBNP ($\geq$353 pg/mL) is best predicted (AUROC $>$ 0.90). Other values with AUROC $>$ 0.85 include Hemoglobin ($\geq$17.5 g/dL), Albumin ($\geq$5.2 g/dL), and Hematocrit ($\geq$51\%). Our findings show ECG combined with clinical data enables prompt abnormality prediction and forecasting of lab abnormalities, offering a non-invasive, cost-effective alternative to traditional testing. This can support early intervention and enhanced patient monitoring. ECG and clinical data can help estimate and monitor abnormal lab values, potentially improving care while reducing reliance on invasive and costly procedures.
\end{abstract}
\begin{document}

\flushbottom
\maketitle

\thispagestyle{empty}

\section{Introduction}

Laboratory values play a crucial role in various medical contexts, ranging from diagnosing diseases\cite{chernecky2012laboratory} to guiding therapeutic interventions in routine care \cite{rigby2017review} and intensive care settings \cite{ezzie2007laboratory}. These values, derived from various bodily fluids or tissues, provide quantitative insights into a  patient's physiological status, aiding clinicians in making informed decisions. Despite their importance, obtaining laboratory values often involves invasive procedures, such as venipuncture or arterial puncture, which can cause discomfort and carry associated risks \cite{ialongo2016phlebotomy,buowari2013complications}. Moreover, the process is resource-intensive \cite{beliveau2018decreasing,halpern2004critical}, requiring specialized equipment, skilled personnel, and significant time for sample collection, processing, and analysis. Delays in obtaining results further limit their utility in rapidly evolving clinical scenarios, highlighting the need for alternative or complementary approaches of measurement \cite{shah2014accuracy}.

An electrocardiogram (ECG) is essential as a first-in-line assessment of a patient's cardiac state. While the ECG is traditionally associated with the detection of cardiovascular diseases, the advent of deep learning has expanded the diagnostic potential of the ECG towards other domains, such as non-cardiac conditions, as reviewed by  Topol et al.\,\cite{topol2021s} and  Siontis et al.\,\cite{siontis2021artificial}, and demonstrated within a single model\cite{strodthoff2024prospects}, or as an important predictor for patient deterioration \cite{alcaraz2024mdsedmultimodaldecisionsupport}. Recent research has explored the relationship between cardiac patterns observed on ECG and abnormal laboratory values, suggesting potential links between systemic health and cardiovascular function, and while advancements in machine learning have enabled ECG to model diverse physiological factors beyond cardiovascular health, the integration of ECG with machine learning towards modeling laboratory values presents new opportunities for learning complex interactions which can enhance clinical decision making \cite{hager2024evaluation}.

\subsection{Background}

\heading{Laboratory values and ECG changes}
Several studies have explored how laboratory values correlate with ECG changes.  Surawicz et al.\,\cite{surawicz1967relationship} classified ECG abnormalities due to electrolyte imbalances, distinguishing between reversible changes from ion shifts and structural damage from chronic deficiencies.  Núñez et al.\,\cite{nunez2009relationship} identified low lymphocyte counts as a prognostic marker for mortality or myocardial infarction in patients with non-diagnostic ECGs.  Abe et al.\,\cite{abe1996electrocardiographic} examined the link between ECG abnormalities and urea nitrogen levels in hemodialysis patients, noting prevalent arrhythmias and ST-T changes.  Shashikala et al.\,\cite{shashikala2014correlation} found significant ECG abnormalities correlated with lower hemoglobin levels in anemic patients.  Laudanski et al.\,\cite{laudanski2009relationship} highlighted a strong correlation between high serum ferritin levels and prolonged QT intervals, indicating potential arrhythmia risks.  Davison et al.\,\cite{davison2024severe} and Trojack et al.\, \cite{trojak2009hypokalemia} found prolonged QTc interval on  hypokalemia. Finally,  Tan et al.\,\cite{tan2021atrial} demonstrated that abnormal ECG patterns interact with biomarkers to influence heart failure risk predictions such as  N-terminal prohormone of brain natriuretic peptide (NT-proBNP).  

\heading{Machine learning and ECG for laboratory values abnormality prediction}
Recent works have focused on predicting laboratory values using ECG data.  Von et al.\,\cite{von2024evaluating} explored predicting potassium, calcium, sodium, and creatinine from ECG signals using deep learning.  Kwon et al.\,\cite{kwon2021artificial} developed a model that classifies electrolyte imbalances such as hyperkalemia and hypokalemia.  Chiu et al.\,\cite{Chiu2024.05.08.24307064} introduced a smartwatch-compatible model for predicting serum potassium levels in end-stage renal disease patients.  Chiu et al.\,\cite{chiu2022utilization} created a personalized model using single-lead ECG data to predict dysglycemia in critically ill patients.  This work stands in the tradition of an earlier conference publication \cite{alcaraz2024cardiolablaboratoryvaluesestimation}, where the feasibility of predicting anomalous laboratory values from ECG features and demographics was demonstrated. Our work extends this by using ECG waveforms with deep learning models and broadening the application from abnormality prediction to both abnormality prediction and abnormality forecasting.

\section{Materials and methods}

\subsection{Clinical workflow and prediction tasks}

\begin{figure*}[ht]
    \centering
    \includegraphics[width=\textwidth]{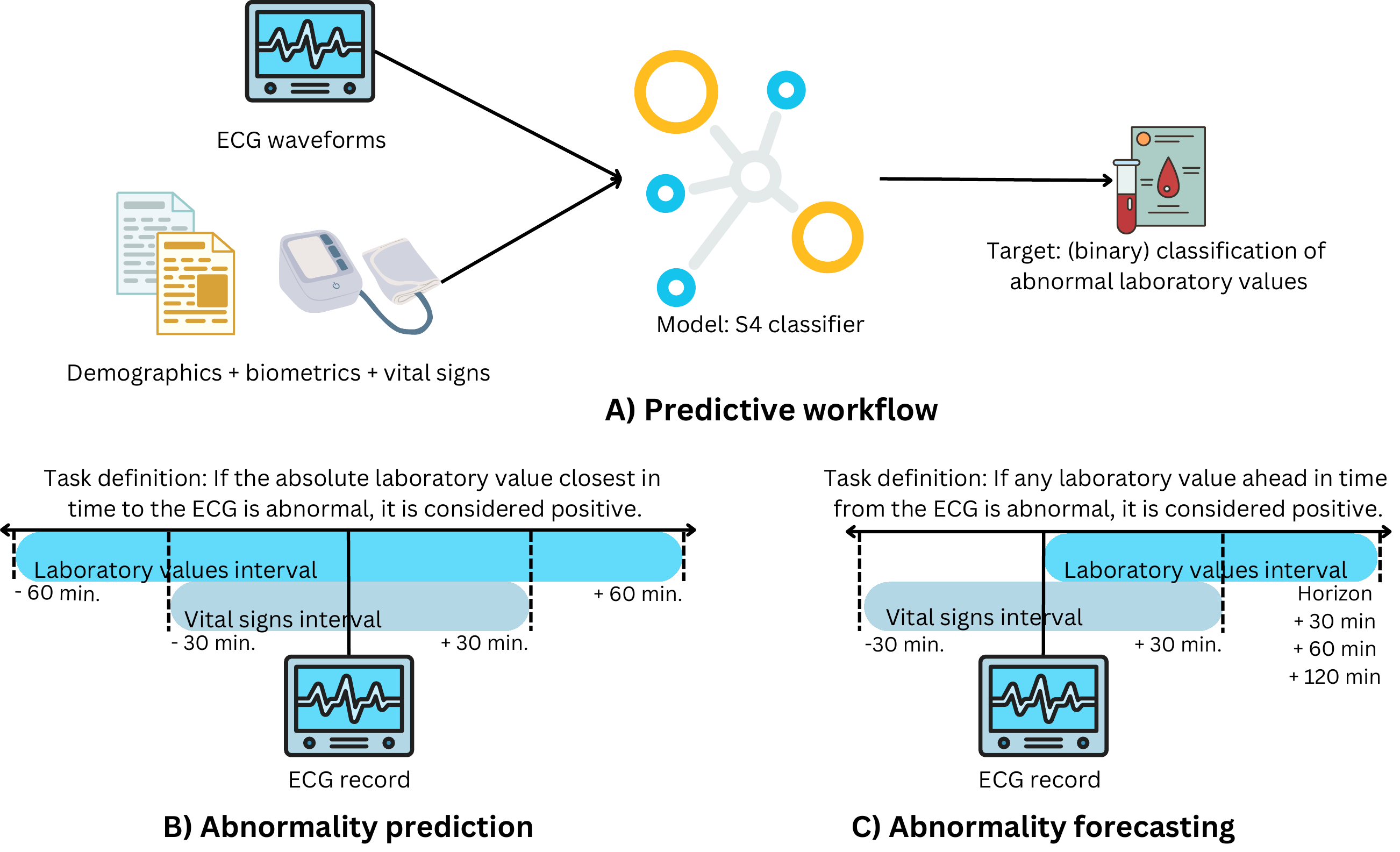}
    \caption{Schematic overview of the study incluiding predictive workflow and task definitions. A) Illustration of the predictive workflow used in the study where model inputs include ECG waveforms, demographics, biometrics, and vital signs, which are used in a binary classification setting to predict abnormal laboratory values. B) Demonstration of the abnormality prediction task, where we sample the closest vital signs within 30 minutes around the ECG record and use the closest laboratory value within 60 minutes as prediction target. C) Demonstration of the abnormality forecasting task, where we also use closest vital signs within 30 minutes around the ECG record as input but where we predict the presence of any abnormal laboratory value within a predefined time horizon.}
    \label{fig:abstract}
\end{figure*}

Figure~\ref{fig:abstract} provides a schematic overview of this study. The illustration of the predictive workflow of CardioLab is shown in Figure~\ref{fig:abstract}, which is composed of a pipeline that combines ECG waveforms and tabular clinical features as inputs to a deep learning model, aiming to predict laboratory value abnormalities as binary classification tasks. The two investigated predictive tasks in this study include abnormality prediction and abnormality forecasting. The abnormality prediction task, shown in Figure~\ref{fig:abstract} B, utilizes vital signs from 30 minutes collected before or after the ECG was taken to predict laboratory abnormalities based on the closest value within 60 minutes. The abnormality forecasting task, depicted in Figure~\ref{fig:abstract} C, uses the same vital signs to forecast if an abnormality occurs within 30, 60 or 120 minutes in the future.  We require at least 10 positive and 10 negative cases per fold, resulting in 137 abnormalities for abnormality prediction, and 116, 126, and 135 abnormalities for abnormality forecasting at time horizons 30, 60, and 120 minutes, respectively.

\subsection{Dataset creation}

\begin{figure*}[ht]
    \centering
    \includegraphics[width=\textwidth]{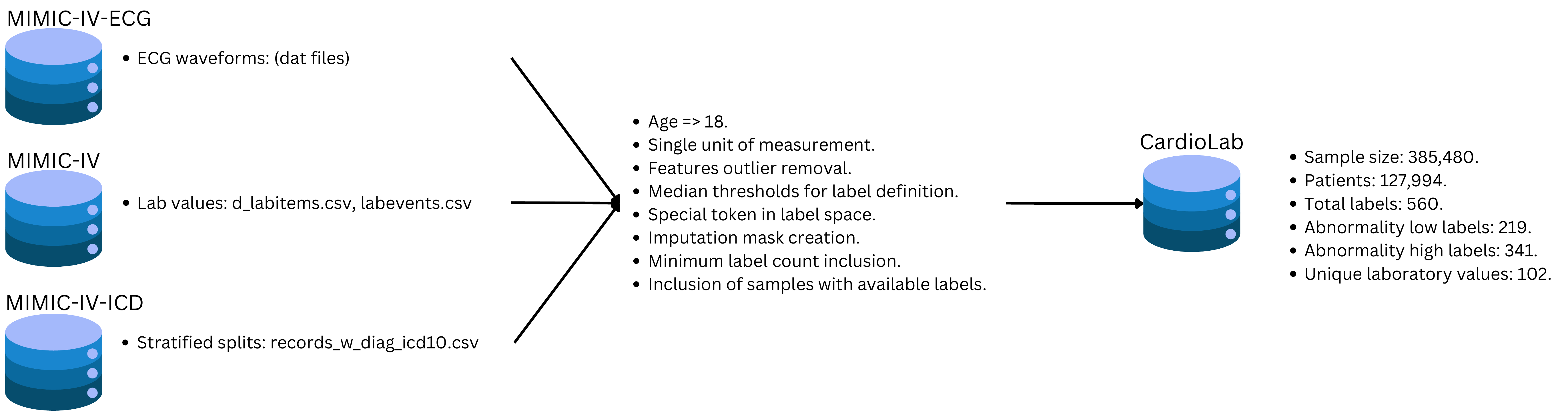}
    \caption{Schematic overview of the dataset creation process, starting from three source datasets (MIMIC-IV-ECG, MIMIC-IV, and MIMIC-IV-ECG-ICD) and applying various inclusion criteria. The process includes patient selection equal or above 18 years old, unit standardization for laboratory values, outlier removal, and label definition based on median-derived thresholds.  Similarly, we incorporate a special token in the label space for labels that were not available to sample for that specific sample, these tokens help as placeholders to be excluded during training and evaluation which ensure large availability of samples and labels even when some labels are not available for a sample. Additionally, we incorporated an imputation strategy to handle missing data with the use of binary masks as helpers e.g. one binary feature per each feature to let the model know which values are real and which imputed. Furthermore, we selected only labels with at least 20 available (10 positive and 10 negative) counts. Finally, we dropped samples that does not contain any available labels after the already mentioned steps.}
    \label{fig:dataset_creation}
\end{figure*}

Figure~\ref{fig:dataset_creation} represents an schematic overview of dataset creation process used in this study. It is based on combining MIMIC-IV-ECG ECG traces with MIMIC-IV clinical data using the stratified splits proposed in MIMIC-IV-ECG-ICD. We restrict ourselves to patients aged 18 years or older. We investigate a single unit of measurement for each laboratory value based on the most prevalent unit of measurement in the dataset. We apply outlier removal, which is primarily error-based, excluding unrealistic values, never-registered extremes, or negative values when the minimum is zero. To define thresholds for abnormally high or low values per laboratory value, we leverage patient-specific thresholds per laboratory value, which are available as part of the clinical data in MIMIC-IV. As the origin of these patient-specific thresholds is not disclosed, we use the median threshold values across the entire dataset as patient-independent threshold values. We introduce a special token to indicate cases where no information on a particular laboratory value is available, next to the binary prediction targets (normal/abnormal). Masking these values during training and evaluation allows to train a single prediction model for the entire set of laboratory values and different prediction horizons.  We considered labels, i.e. abnormally low or abnormally high laboratory values, which at this stage contain at least 10 counts in validation and test set independently. Finally, we include only the samples that after all the previously mentioned criteria contain at least one available label across all investigated horizons.

The final dataset consists of 385,480 samples from 127,994 patients, covering 560 distinct labels derived from 102 unique laboratory values. Among these labels, 341 indicate abnormally high values, while 219 correspond to abnormally low values. Although the MIMIC-IV dataset primarily represents a US ICU/ED population and may not be fully representative of all demographic groups, it remains the most extensive and publicly available critical care dataset. Its broad coverage of patient records across diverse demographics and thousands of multimodal features enables a comprehensive predictive modeling.

\subsection{Features}

Each sample features a 12-lead ECG recorded for exactly 10 seconds. Demographic data includes gender, age, and race (Caucasian, African, Asian, Latino, or other). We consider widely available vital signs such as temperature (°F), heart rate (bpm), respiration rate (bpm), oxygen saturation (\%), systolic and diastolic blood pressure (mmHg). Biometrics such as body mass index (BMI) (kg/m$^2$), weight (kg), and height (cm) are also included. For the tabular features entering the multimodal model, we apply median imputation using training set statistics and introduce as binary column to indicate if a particular value was imputed, see  Alcaraz et al.\,\cite{alcaraz2024mdsedmultimodaldecisionsupport} for details on the effectiveness of this procedure.

\subsection{Train-test splits} 

We utilize the stratified splits provided in  Strodthoff et al.\,\cite{strodthoff2024prospects}, ensuring balanced distribution across gender, age groups, and discharge diagnoses. The dataset is divided into 20 stratified folds, allocated to training, validation, and testing in a 18:1:1 ratio.

\subsection{Models}

The primary model used is an S4 classifier, featuring a time series encoder composed of four bidirectional S4 blocks \cite{Gu2021EfficientlyML}, with hyperparameter configurations adopted from prior work \cite{mehari2023towards}. Specifically, each S4 block has a dimension of 512 and a state size of 8. This architecture has been successfully applied in previous ECG classification studies \cite{mehari2023towards,strodthoff2024prospects,alcaraz2024mdsedmultimodaldecisionsupport}, demonstrating strong performance  and outperforming widely used baselines such as ResNets in physiological time series tasks \cite{strodthoff2024prospects}. Given the similarity in waveform data applications, no modifications were required. Additionally,  Strodthoff et al.\,\cite{strodthoff2024prospects} provide a benchmarking study comparing alternative deep learning architectures, reaffirming the superiority of the S4 classifier over modern convolutional, recurrent, and transformer-based models for signal encoding. As a tabular encoder for static features we use a three-layer  multi-layer perceptron (MLP), where categorical features (gender and ethnicity) are first processed through embedding layers. The outputs of the signal encoder and tabular encoder are then fused via concatenation.

\subsection{Training}

For the S4 models, we used AdamW \cite{loshchilov2018decoupled} as optimizer, setting both the learning rate and weight decay to 0.001 while using a constant learning rate schedule. Training was performed with a batch size of 64 samples for 20 epochs. These hyperparameter choices align with those from a prior unimodal study \cite{strodthoff2024prospects}, reinforcing the reliability of our results without extensive hyperparameter tuning. To prevent overfitting, we applied model selection based on validation performance, i.e., after training the best-performing model according to validation set performance was selected for final evaluation, rendering the total epoch count inconsequential as long as it was sufficient for convergence. The model was trained to minimize binary cross-entropy loss. Following  Mehari et al.\,\cite{mehari2023towards}, we use 2.5-second ECG segments  in contrast to the available 10-seconds segments, as longer windows substantially increase computational complexity without providing significant improvement in predictive performance. During testing, predictions are averaged over four non-overlapping 2.5-second segments.  Similarly, the original ECG signals in MIMIC-IV-ECG are recorded at 500 Hz, but also following Mehari et al.\,\cite{mehari2023towards} we downsample to 100 Hz for computational efficiency and consistency with prior work on diagnostic tasks. Training takes about 19 hours on a NVIDIA L40 GPU.  Other requirements include Python 3.10.8, PyTorch 1.13.0, and PyTorch Lightning 1.8.0.

\subsection{Performance evaluation}

Statistical uncertainty is assessed using empirical bootstrap with 1000 iterations, providing 95\% prediction intervals for macro area under the receiver operating characteristic curve (AUROC) and individual AUROCs. We choose macro AUROC as the primary evaluation metric because it is the most widely used ranking-based measure. It assesses the model's overall discriminative ability without requiring predefined decision thresholds. Additionally, recent theoretical and empirical research \cite{mcdermott2024closer} has validated AUROC's suitability over other common metrics, such as the area under the precision-recall curve (AUPRC), particularly in the presence of label imbalance.

Finally, although we primarily assess predictive performance in the main manuscript using AUROC, the supplementary material also reports additional metrics. Specifically, we provide sensitivity and specificity at a fixed operating point (sensitivity = 0.75) for all labels in Supplementary Tables S1 and S2, corresponding to abnormality prediction and forecasting, respectively. We emphasize that this operating point is an arbitrary example, and thresholds should be chosen label-wise based on clinical considerations for deployment. In addition, we report label prevalence, defined as the proportion of positive cases over the total sample size, together with the corresponding sample counts in Supplementary Tables S3 and S4, again for abnormality prediction and forecasting, respectively.

\section{Results}

\subsection{Descriptive statistics}

\begin{table}[ht]
\centering
\caption{Summary statistics across samples. For demographics it covers gender distribution, age distribution (median
with standard deviation and quartiles), ethnicity distribution in the study population, presented as percentages. For vital signs it shows temperature, heart rate, respiration rate, oxygen saturation, systolic, and diastolic blood pressure median and quantiles and finally  body mass index, height, and weight median and quantiles for biometrics.}
\begin{tabular}{l c} 
\hline\hline
\textbf{Name}  & \textbf{Values}  \\ 
\hline
\textbf{Gender} &  \\
\quad Male & 194,343 (50.41\%)\\
\quad Female & 191,137 (49.58\%) \\
\textbf{Age \%} &  \\
\quad Median years (IQR) & 64 (25) \\
\quad Quantile 1: 18-50 &  97,606 (25.32\%) \\
\quad Quantile 2: 51-63 &  102,045 (26.47\%) \\
\quad Quantile 3: 64-75 &  95,903 (24.87\%) \\
\quad Quantile 4: 76-91 &  89,926 (23.32\%) \\
\textbf{Ethnicity \%} &  \\
\quad Caucasian & 251,872 (65.33\%) \\
\quad Asian &  11,795 (3.05\%) \\
\quad African & 64,733 (16.79\%) \\
\quad Hispanic & 22,509 (5.83\%) \\
\quad Other & 28,633 (7.42\%) \\
\textbf{Vital signs} &  \\
\quad Temperature &  98 (1) \\
\quad Heart rate &  83 (29) \\
\quad Respiration rate &  18 (4) \\
\quad Oxygen saturation & 99 (3) \\
\quad Systolic blood pressure & 132 (34) \\
\quad Diastolic blood pressure & 74 (22) \\
\textbf{Biometrics} &  \\
\quad BMI & 27 (9) \\
\quad Height & 167 (15) \\
\quad Weight & 77 (27) \\
\hline\hline
\end{tabular}
\label{table:descriptive}
\end{table}

Table \ref{table:descriptive} summarizes the characteristics of the study population. Gender distribution is nearly balanced (50.41\% male, 49.58\% female). The median age is 64 years (IQR: 25), with age quantiles spanning 18-91 years. Ethnicity is predominantly Caucasian (65.33\%), followed by African (16.79\%), Hispanic (5.83\%), Asian (3.05\%), and Other (7.42\%). Vital signs include a median temperature of 98°F, heart rate of 83 bpm, respiration rate of 18 breaths/min, Oxygen saturation of 99\%, systolic blood pressure of 132 mmHg, and diastolic blood pressure of 74 mmHg. Biometrics show a median BMI of 27 kg/m$^2$, height of 167 cm, and weight of 77 kg.

\subsection{Abnormality prediction}

\begin{table*}[ht]
\caption{Performance results for laboratory values abnormality prediction with AUROC exceeding 0.7 (i.e., lower bound of the AUROC score larger than 0.7), presented by value name, threshold and nature of the abnormality, and corresponding AUROC scores with 95\% prediction intervals. Next to each value name, we highlight the physiological category it represents: Cardiac (Ca.), Renal (Re.), Hematological (He.), Metabolic (Me.), Immunological (Im.), and Coagulation (Co.).}
\centering
\begin{tabular}{lccc} \hline\hline
\textbf{Value}  & \textbf{Threshold}  & \textbf{Abnormality prediction AUROC} \\ \hline
NTproBNP [Ca.] & $\geq$353 pg/mL  & 0.903 (0.883, 0.923) \\
Hemoglobin [He.] & $\geq$17.5 g/dL & 0.870 (0.816, 0.919) \\
Albumin [Me.] & $\geq$5.2 g/dL & 0.859 (0.744, 0.953) \\
Acetaminophen [Me.] & $\geq$30 ug/mL & 0.846 (0.750, 0.931) \\
Hematocrit [He.] & $\geq$51\% & 0.821 (0.768, 0.873) \\
Red Blood Cells [He.] & $\geq$6.1 m/uL & 0.791 (0.718, 0.850) \\
Urea Nitrogen [Re.] & $\leq$6 mg/dL & 0.790 (0.747, 0.830) \\
Creatinine [Re./Ca.] & $\geq$1.2 mg/dL & 0.788 (0.780, 0.797) \\
Bilirubin, Direct [Me.] & $\geq$0.3 mg/dL & 0.786 (0.716, 0.852) \\
Urea Nitrogen [Re.] & $\geq$20 mg/dL & 0.786 (0.778, 0.795) \\
Albumin [Me.] & $\leq$3.5 g/dL & 0.771 (0.751, 0.790) \\
Hemoglobin [He.] & $\leq$13.7 g/dL & 0.771 (0.761, 0.780) \\
Cholesterol HDL [Me.] & $\leq$41 mg/dL & 0.765 (0.716, 0.814) \\
RDW-SD [He.] & $\geq$46.3 fL & 0.764 (0.750, 0.779) \\
Absolute Basophil Count [Im.] & $\leq$0.01 K/uL & 0.746 (0.711, 0.781) \\
Creatinine [Re./Ca.] & $\leq$0.5 mg/dL & 0.746 (0.707, 0.786) \\
Red Blood Cells [He.] & $\leq$4.6 m/uL & 0.744 (0.734, 0.754) \\
RDW [He.] & $\geq$15.5\% & 0.743 (0.732, 0.753) \\
Hematocrit [He.] & $\leq$40\% & 0.742 (0.732, 0.751) \\
INR PT [Co.] & $\geq$1.1 ratio & 0.736 (0.726, 0.748) \\
Bilirubin, Total [Me.] & $\geq$1.5 mg/dL & 0.733 (0.706, 0.761) \\
PT [Co.] & $\geq$12.5 sec & 0.730 (0.720, 0.742) \\
Lymphocytes [Im.] & $\leq$18\% & 0.719 (0.708, 0.730) \\
MCHC [He.] & $\leq$31 g/dL & 0.718 (0.701, 0.735)  \\ \hline\hline
\end{tabular}
\label{table:new_est_main}
\end{table*}

Table \ref{table:new_est_main} presents the performance results for laboratory value abnormality prediction, highlighting 24 abnormalities across 19 unique laboratory values that achieved statistically significant AUROC scores above 0.7. Notably, NTproBNP levels above 353 pg/mL were estimated with an AUROC of 0.903. Urea nitrogen levels below 6 mg/dL were estimated with an AUROC of 0.790. Hemoglobin levels above 17.5 g/dL had an AUROC of 0.870, while Creatinine levels below 0.5 mg/dL had an AUROC of 0.746. Other values, such as bilirubin, hematocrit, and albumin, also demonstrated strong predictive performance.

\subsection{Abnormality forecasting}

\begin{table*}[ht]
\caption{Performance results for laboratory values abnormality forecasting with AUROC exceeding 0.7 (i.e., lower bound of the AUROC score larger than 0.7 in at least one setting), presented by value name, threshold and nature of the abnormality, and corresponding AUROC scores with 95\% prediction intervals for the 30-minute, 60-minute, and 120-minute horizons. The sorting criteria for the table are defined by the mean of test AUROC across the three horizons. Next to each value name, we highlight the physiological category it represents: Cardiac (Ca.), Renal (Re.), Hematological (He.), Metabolic (Me.), Immunological (Im.), and Coagulation (Co.). The values in bold represent the highest performance per abnormality. For PT at the 30~minute horizon no score was estimated as it did not satisfy the inclusion criterion of at least 10 abnormal cases in both test and validation set.}
\centering
\begin{tabular}{lccccc} \hline\hline
\textbf{Value}  & \textbf{Threshold}  & \textbf{30 min. AUROC} & \textbf{60 min. AUROC} & \textbf{120 min. AUROC} \\ \hline
NTproBNP [Ca.] & $\geq$353 pg/mL & 0.903 (0.87, 0.933) & 0.907 (0.883, 0.929) & \textbf{0.916 (0.898, 0.934)}\\
Hemoglobin [He.]& $\geq$17.5 g/dL  & 0.903 (0.841, 0.96) & \textbf{0.911 (0.867, 0.951)} & 0.90 (0.856, 0.938) \\
Hematocrit [He.] & $\geq$51\% & \textbf{0.85 (0.781, 0.914)} & 0.844 (0.79, 0.896) & 0.843 (0.794, 0.896) \\
Urea Nitrogen [Re.]& $\leq$6 mg/dL & 0.793 (0.717, 0.859) & 0.813 (0.761, 0.862) & \textbf{0.84 (0.80, 0.877)} \\
Creatinine [Re./Ca.] & $\geq$1.2 mg/dL & 0.792 (0.779, 0.805) & 0.801 (0.79, 0.811) & \textbf{0.805 (0.796, 0.814)} \\
Urea Nitrogen [Re.]& $\geq$20 mg/dL  & 0.795 (0.783, 0.808) & 0.798 (0.788, 0.808) & \textbf{0.799 (0.79, 0.807)} \\
Red Blood Cells [He.] & $\geq$6.1 m/uL  & \textbf{0.813 (0.744, 0.876)} & 0.784 (0.702, 0.854) & 0.774 (0.695, 0.843) \\
Hemoglobin [He.]& $\leq$13.7 g/dL & 0.781 (0.767, 0.794) & 0.779 (0.767, 0.79) & \textbf{0.781 (0.77, 0.791)} \\
Albumin [Me.] & $\leq$3.5 g/dL & 0.769 (0.738, 0.798) & \textbf{0.778 (0.753, 0.801)} & 0.778 (0.755, 0.799) \\
RDW-SD [He.] & $\geq$ 46.3 fL & 0.768 (0.745, 0.792) & 0.776 (0.757, 0.794) & \textbf{0.773 (0.759, 0.788)} \\
Absolute Monocyte Count [Im.] & $\leq$0.2 K/uL & 0.744 (0.663, 0.821) & \textbf{0.769 (0.711, 0.827)} & 0.75 (0.698, 0.803) \\
RDW [He.]& $\geq$15.5\%   & 0.75 (0.734, 0.766) & 0.755 (0.742, 0.768) & \textbf{0.756 (0.744, 0.766)} \\
Absolute Lymphocyte Count [Im.] & $\geq$ 3.7 K/uL & \textbf{0.777 (0.69, 0.866)} & 0.768 (0.707, 0.826) & 0.714 (0.648, 0.781) \\
Red Blood Cells [He.] & $\leq$4.6 m/uL & \textbf{0.757 (0.742, 0.771)} & 0.75 (0.737, 0.762) & 0.748 (0.737, 0.759) \\
C-Reactive Protein [Im.] & $\geq$ 5 mg/L & 0.714 (0.592, 0.833) & 0.754 (0.657, 0.844) & \textbf{0.783 (0.703, 0.852)} \\
Hematocrit [He.]& $\leq$40\% & \textbf{0.752 (0.738, 0.764)} & 0.748 (0.735, 0.759) & 0.746 (0.736, 0.757) \\
Absolute Basophil Count [Im.] & $\leq$0.01 K/uL & 0.722 (0.664, 0.779) & \textbf{0.754 (0.71, 0.796)} & 0.752 (0.71, 0.789) \\
INR PT [Co.] & $\geq$1.1 ratio  & \textbf{0.74 (0.725, 0.757)} & 0.74 (0.726, 0.755) & 0.74 (0.729, 0.753) \\
Lymphocytes [Im.]& $\geq$42\%  & \textbf{0.735 (0.7, 0.768)} & 0.735 (0.705, 0.762) & 0.735 (0.71, 0.759) \\
PT [Co.]& $\geq$12.5 sec & \textbf{0.735 (0.719, 0.751)} & 0.735 (0.721, 0.749) & 0.733 (0.721, 0.745) \\
Lymphocytes [Im.]& $\leq$18\% & \textbf{0.735 (0.719, 0.75)} & 0.725 (0.712, 0.738) & 0.731 (0.72, 0.743) \\
Anion Gap [Me.] & $\geq$20 mEq/L  & \textbf{0.741 (0.715, 0.765)} & 0.726 (0.704, 0.75) & 0.721 (0.701, 0.74) \\
MCHC [He.]& $\leq$31 g/dL & \textbf{0.733 (0.708, 0.757)} & 0.728 (0.707, 0.747) & 0.719 (0.702, 0.736) \\
Bilirubin Total [Me.] & $\geq$1.5 mg/dL & 0.708 (0.659, 0.753) & 0.723 (0.686, 0.761) & \textbf{0.737 (0.706, 0.768)} \\ \hline\hline
\end{tabular}
\label{table:mon_main}
\end{table*}

Table \ref{table:mon_main} presents the AUROC performance results for laboratory value abnormality forecasting across different abnormality thresholds and physiological categories. The results highlight the predictive capacity of various laboratory values over three forecasting horizons (30, 60, and 120 minutes). Overall, 24 distinct abnormalities and 21 unique laboratory values exhibited a statistically significant lower bound AUROC exceeding 0.7. Among the most predictive markers, NTproBNP as a cardiac marker with levels above 353 pg/mL demonstrated the highest performance. Similarly, elevated hemoglobin levels ($\geq$ 17.5 g/dL) and hematocrit levels ($\geq$ 51\%) showed strong predictive capabilities, with AUROC values exceeding 0.84 across all horizons, highlighting their relevance in hematological assessments. Renal function markers such as urea nitrogen and creatinine also showed promising predictive performance. Low urea nitrogen levels ($\leq$ 6 mg/dL) demonstrated an AUROC of 0.84 at 120 minutes, while high creatinine levels ($\geq$ 1.2 mg/dL) achieved an AUROC of 0.805 at the same horizon. These findings align with the clinical significance of these markers in detecting early renal dysfunction and associated complications. Markers related to coagulation and immune response, such as PT, and absolute monocyte counts, exhibited moderate predictive performance. For instance, PT values above 12.5 seconds showed a significant AUROC of 0.733 at 60 minutes, while absolute monocyte counts below 0.2 K/uL had an AUROC of 0.769 at 30 minutes. The role of these parameters in inflammatory and coagulation processes is well-documented, supporting their relevance in early abnormality detection. The predictive utility of metabolic markers such as albumin and bilirubin was also evident. Low albumin levels ($\leq$ 3.5 g/dL) yielded an AUROC of 0.778 at 60 minutes, reinforcing its importance in nutritional and metabolic assessments. Additionally, elevated total bilirubin levels ($\geq$ 1.5 mg/dL) achieved an AUROC of 0.737 at 120 minutes, suggesting its potential in abnormality forecasting hepatic function.

\section{Discussion}

\heading{Summary}
This study explores the potential of using ECG data combined with demographic and clinical data to estimate and monitor laboratory value abnormalities. Utilizing the MIMIC-IV dataset, multimodal deep-learning models were developed to predict  prompt abnormalities and at future intervals. The models achieved strong performance, with AUROC scores above 0.70 (lower bound) for up to 24 laboratory values across various physiological categories in the abnormality prediction setting and 24 in the abnormality forecasting ones. Key results include accurate predictions for NTproBNP, hemoglobin, hematocrit, urea nitrogen, and creatinine levels. These findings demonstrate the feasibility of non-invasive, cost-effective methods for patient monitoring, enhancing the way for diagnostic tools and early interventions in clinical settings.

\heading{Physiological groups and their relationship with ECG Data}
The ability to predict various laboratory abnormalities from ECG data is clinically significant as it reflects underlying physiological changes. Cardiac markers such as NTproBNP, creatinine, and creatine kinase are directly related to heart stress and injury, with ECG changes providing insight into these conditions. Renal markers like urea nitrogen, creatinine, and calcium can cause cardiac function disturbances due to fluid and electrolyte imbalances, making ECG prediction plausible \cite{boudoulas2017cardio}. Hematological values, including hemoglobin, hematocrit, and RDW, are linked to blood disorders affecting cardiac function, which can be mirrored in ECG changes. Metabolic markers such as bicarbonate and anion gap influence cardiac function through disturbances like acidosis or alkalosis, detectable by ECG. Immunological parameters like iron binding and lymphocytes may indirectly affect cardiac health through systemic inflammation, which can be reflected in ECG patterns \cite{swirski2018cardioimmunology}. Finally, coagulation markers like INR PT and PT are essential for assessing bleeding and clotting risks, with their abnormalities potentially causing ECG changes due to associated cardiac stress. 

\heading{Clinical relevance and applications} 
The model's ability to predict abnormal laboratory values in the abnormality prediction not only for laboratory values commonly associated with cardiac diseases mirrors the ability to infer hints on non-cardiac conditions from the ECG \cite{ceasovschih2024electrocardiogram}. The abnormality forecasting aligns with prior work highlighting the added diagnostic value of ECG waveforms for deterioration task \cite{alcaraz2024mdsedmultimodaldecisionsupport}. Nevertheless, the prediction of anomalous laboratory values represents a prediction task with a well-defined ground truth which makes it a relevant case for algorithmic benchmarking. Clinical implications of predicting laboratory values from ECG data, such as potential benefits in patient risk identification \cite{oloyede2024identifying}, patient monitoring for adverse drug events \cite{oloyede2025predicting}, enhanced performance for early diagnosis \cite{alcaraz2024explainable}, treatment personalization, and enhanced performance for faster and low-cost diagnoses 
\cite{alcaraz2025electrocardiogram}. As an example, predicting electrolyte imbalances (such as potassium and calcium) from ECG data can enable early detection of arrhythmias, allowing timely interventions in emergency or ICU settings. Additionally, predicting NTproBNP levels can aid in the early diagnosis and monitoring of heart failure, especially in remote monitoring or routine check-ups, enabling preventive action \cite{chami2022point}. Creatinine predictions from ECG could provide early insights into renal function, aiding in the detection of kidney dysfunction in outpatient or home care settings, preventing further damage through earlier referrals.

\heading{Limitations} Our approach has several limitations. First, our task definition relies on median threshold values inferred from MIMIC-IV and therefore do not cover dependencies of reference values on patient characteristics such as demographics. This question will have to be revisited more thoroughly in future studies. For certain scenarios, also the differentiation into anomalously low and anomalously high values might be not finegrained enough and might necessitate to frame the prediction problem as a multi-class ordinal classification task or even as a regression problem. Second, the current setup of the abnormality forecasting task does not provide any insights into the time frame where the abnormality occurs. This could be mitigated by predicting future laboratory values at a particular future horizon up to some tolerance. Here, we mainly use it to demonstrate the feasibility of providing prognostic insights into laboratory values. Third, the approach should be tested on diverse cohorts to ensure it generalizes effectively before being deployed in clinical settings.

\heading{Future work} 
First, we believe that a natural next step after the demonstration of the general feasibility of predicting and forecasting abnormalities across a broad variety of laboratory values would be to focus on individual laboratory values or groups thereof more closely, incorporating state-of-the-art reference intervals, more adapted/finegrained prediction scenarios and external validation in appropriate cohorts. Second, we envision the use of explainable AI techniques, such as those put forward by  Wagner et al.\,\cite{wagner2023explaining} or concept-based causal approaches \cite{alcaraz2024causalconceptts}, which could enhance our understanding of the mechanisms of how abnormal laboratory values impact the ECG. Third, establishing population-specific baselines (e.g., across age, sex, or ethnicity) could provide valuable context to better differentiate between normal and pathological signal changes in future applications \cite{ott2024using}.  Fourth, evaluating prompt abnormality prediction and forecasting in real clinical settings, while accounting for latency, integration with continuous vital sign monitoring, and computational overhead, would yield important insights into feasibility and robustness. Importantly, very low latency is not required; even a delay of a few minutes is acceptable and still far superior to the current gold standard of waiting for lab results. Fifth, evaluating cases where laboratory values transition from normal to abnormal within forecasting horizons could provide deeper insights into the temporal evolution of patient physiology and the robustness of predictive models. Finally, for clinical deployment, one could investigate evaluation schemes at specific operating points, such as sensitivity and specificity thresholds, to better understand false positive and false negative trade-offs however clinical context should be included and the analysis should focus on a smaller set of laboratory values.

\bibliography{sample}
\clearpage

\heading{Author contribution statement} JMLA and NS were responsible for the conceptualization of the project. Data curation was performed by JMLA and NS. JMLA and NS implemented the prediction models, with JMLA conducting the formal analysis under NS's supervision. JMLA produced the original draft, and NS contributed to the review and editing. All authors participated in the investigation, methodology, and validation, and they approved the final version for publication. Computing resources were provided through Carl von Ossietzky University of Oldenburg. All figures in this manuscript were created by the authors, and no generative AI tools were used for figure creation.

\heading{Declaration of generative AI and AI-assisted technologies in the writing process} During the preparation of this work, the authors used ChatGPT (OpenAI) in order to refine language and correct grammar. After using this tool, the authors reviewed and edited the content as needed and take full responsibility for the content of the publication.

\heading{Data and code availability} Code for dataset preprocessing and experimental replications can be found in our dedicated repository \url{https://github.com/AI4HealthUOL/CardioLab}.

\heading{Hardware and software requirements} The experiments were conducted on single NVIDIA L40 GPUs. The environment included Python 3.10.8, PyTorch 1.13.0, and PyTorch Lightning 1.8.0 to perform the computational tasks.

\heading{Conflicts of interest} 
Upon manuscript submission, all authors completed the author disclosure form, confirming the absence of any conflicts of interest.

\clearpage

\appendices

\section*{Predictive performance at operating point}

\begin{table*}[ht]
\caption{Sensitivity (sens.) and specificity (spec.) at sensitivity theshold 0.75 for the labels investigated in the main manuscript across the task laboratory values abnormality prediction. Next to each value name, we highlight the physiological category it represents: Cardiac (Ca.), Renal (Re.), Hematological (He.), Metabolic (Me.), Immunological (Im.), and Coagulation (Co.). The values in bold represent the highest performance per abnormality. For PT at the 30~minute horizon no score was estimated as it did not satisfy the inclusion criterion of at least 10 abnormal cases in both test and validation set.}
\centering
\begin{tabular}{lcc} \hline\hline
\textbf{Value}  & \textbf{Threshold}  & \textbf{Sensitivity/Specificity} \\ \hline
NTproBNP [Ca.] & $\geq$353 pg/mL  &  0.751/0.881  \\
Hemoglobin [He.] & $\geq$17.5 g/dL  &  0.750/0.822   \\
Albumin [Me.] & $\geq$5.2 g/dL & 0.750/0.742  \\
Acetaminophen [Me.] & $\geq$30 ug/mL  & 0.768/0.839 \\
Hematocrit [He.] & $\geq$51\%  & 0.750/0.761  \\
Red Blood Cells [He.] & $\geq$6.1 m/uL  &  0.750/0.699  \\
Urea Nitrogen [Re.] & $\leq$6 mg/dL  & 0.750/0.696  \\
Creatinine [Re./Ca.] & $\geq$1.2 mg/dL  &  0.750/0.675  \\
Bilirubin, Direct [Me.] & $\geq$0.3 mg/dL  &  0.752/0.638  \\
Urea Nitrogen [Re.] & $\geq$20 mg/dL  & 0.750/0.674 \\
Albumin [Me.] & $\leq$3.5 g/dL  & 0.750/0.642   \\
Hemoglobin [He.] & $\leq$13.7 g/dL  & 0.750/0.651 \\
Cholesterol HDL [Me.] & $\leq$41 mg/dL  & 0.755/0.676 \\
RDW-SD [He.] & $\geq$46.3 fL  & 0.751/0.654  \\
Absolute Basophil Count [Im.] & $\leq$0.01 K/uL & 0.750/0.618 \\
Creatinine [Re./Ca.] & $\leq$0.5 mg/dL  & 0.750/0.657 \\
Red Blood Cells [He.] & $\leq$4.6 m/uL  &  0.750/0.606  \\
RDW [He.] & $\geq$15.5\%  &  0.750/0.609 \\
Hematocrit [He.] & $\leq$40\%  &  0.750/0.651 \\
INR PT [Co.] & $\geq$1.1 ratio  &  0.750/0.583 \\
Bilirubin, Total [Me.] & $\geq$1.5 mg/dL  & 0.750/0.581 \\
PT [Co.] & $\geq$12.5 sec & 0.750/0.564  \\
Lymphocytes [Im.] & $\leq$18\%  &  0.750/0.564 \\
MCHC [He.] & $\leq$31 g/dL & 0.750/0.562 \\ \hline\hline
\end{tabular}
\label{table:est_sens_sup}
\end{table*}

\begin{table*}[ht]
\caption{Sensitivity (Sens.) and specificity (Spec.) at sensitivity theshold 0.8 for the labels investigated in the main manuscript across the task laboratory values abnormality prediction. Next to each value name, we highlight the physiological category it represents: Cardiac (Ca.), Renal (Re.), Hematological (He.), Metabolic (Me.), Immunological (Im.), and Coagulation (Co.). The values in bold represent the highest performance per abnormality. For PT at the 30~minute horizon no score was estimated as it did not satisfy the inclusion criterion of at least 10 abnormal cases in both test and validation set.}
\centering
\begin{tabular}{lccccc} \hline\hline
\textbf{Value}  & \textbf{Threshold}  & \textbf{Sens./Spec.} & \textbf{Sens./Spec.} & \textbf{Sens./Spec.} \\ \hline
NTproBNP [Ca.] & $\geq$353 pg/mL & 0.753/0.860 & 0.751/0.884 & 0.751/0.907  \\
Hemoglobin [He.]& $\geq$17.5 g/dL  & 0.750/0.828 & 0.750/0.886 & 0.750/0.815 \\
Hematocrit [He.] & $\geq$51\% & 0.750/0.768 & 0.750/0.744 & 0.755/0.719 \\
Urea Nitrogen [Re.]& $\leq$6 mg/dL & 0.750/0.696 & 0.750/0.747 & 0.750/0.789 \\
Creatinine [Re./Ca.] & $\geq$1.2 mg/dL & 0.750/0.692 & 0.750/0.705 & 0.750/0.709 \\
Urea Nitrogen [Re.] & $\geq$20 mg/dL & 0.750/0.698 & 0.750/0.701 & 0.750/0.701 \\
Red Blood Cells [He.] & $\geq$6.1 m/uL & 0.750/0.669 & 0.750/0.636 & 0.750/0.572 \\
Hemoglobin [He.] & $\leq$13.7 g/dL & 0.750/0.667 & 0.750/0.662 & 0.750/0.666 \\
Albumin [Me.] & $\leq$3.5 g/dL & 0.750/0.650 & 0.750/0.664 & 0.750/0.664 \\
RDW-SD [He.] & $\geq$46.3 fL & 0.751/0.652 & 0.751/0.674 & 0.750/0.648 \\
Absolute Monocyte Count [Im.] & $\leq$0.2 K/uL & 0.750/0.623 & 0.750/0.667 & 0.750/0.626 \\
RDW [He.] & $\geq$15.5\% & 0.750/0.624 & 0.750/0.631 & 0.750/0.631 \\
Absolute Lymphocyte Count [Im.] & $\geq$3.7 K/uL & 0.750/0.601 & 0.750/0.653 & 0.750/0.548 \\
Red Blood Cells [He.] & $\leq$4.6 m/uL & 0.750/0.636 & 0.750/0.619 & 0.750/0.612 \\
C-Reactive Protein [Im.] & $\geq$5 mg/L & 0.756/0.520 & 0.750/0.618 & 0.750/0.673 \\
Hematocrit [He.] & $\leq$40\% & 0.750/0.635 & 0.750/0.622 & 0.750/0.612 \\
Absolute Basophil Count [Im.] & $\leq$0.01 K/uL & 0.75/0.547 & 0.75/0.649 & 0.75/0.630 \\
INR PT [Co.] & $\geq$1.1 ratio & 0.751/0.576 & 0.750/0.587 & 0.750/0.585 \\
Lymphocytes [Im.] & $\geq$42\% & 0.750/0.617 & 0.750/0.600 & 0.750/0.603 \\
PT [Co.] & $\geq$12.5 sec & 0.750/0.567 & 0.750/0.570 & 0.750/0.567 \\
Lymphocytes [Im.] & $\leq$18\% & 0.750/0.600 & 0.750/0.578 & 0.750/0.588 \\
Anion Gap [Me.] & $\geq$20 mEq/L & 0.750/0.611 & 0.750/0.578 & 0.750/0.568 \\
MCHC [He.] & $\leq$31 g/dL & 0.750/0.595 & 0.750/0.590 & 0.750/0.577 \\
Bilirubin Total [Me.] & $\geq$1.5 mg/dL & 0.750/0.540 & 0.750/0.569 & 0.750/0.590 \\ \hline\hline
\end{tabular}
\label{table:mon_sens_sup}
\end{table*}

\clearpage

\section*{Label prevalence and sample count}

\begin{table*}[ht]
\caption{Label prevalence and sample count for the labels investigated in the main manuscript across the tasks laboratory values abnormality prediction. Next to each value name, we highlight the physiological category it represents: Cardiac (Ca.), Renal (Re.), Hematological (He.), Metabolic (Me.), Immunological (Im.), and Coagulation (Co.). The values in bold represent the highest performance per abnormality. For PT at the 30~minute horizon no score was estimated as it did not satisfy the inclusion criterion of at least 10 abnormal cases in both test and validation set.}
\centering
\begin{tabular}{lccc} \hline\hline
\textbf{Value}  & \textbf{Threshold}  & \textbf{Label prevalence}  & \textbf{Sample count}  \\ \hline
NTproBNP [Ca.] & $\geq$353 pg/mL  & 77.72\% & 23,206 \\
Hemoglobin [He.] & $\geq$17.5 g/dL  & 72.39\% & 236,802   \\
Albumin [Me.] & $\geq$5.2 g/dL & 0.4\% & 57,040 \\
Acetaminophen [Me.] & $\geq$30 ug/mL  & 18.76\% & 1,205 \\
Hematocrit [He.] & $\geq$51\%  & 0.58\% & 238,567 \\
Red Blood Cells [He.] & $\geq$6.1 m/uL  & 0.31\% & 234,079 \\
Urea Nitrogen [Re.] & $\leq$6 mg/dL  & 0.89\% & 240,892 \\
Creatinine [Re./Ca.] & $\geq$1.2 mg/dL  & 28.52\% & 241,968 \\
Bilirubin, Direct [Me.] & $\geq$0.3 mg/dL  & 57.24\%  & 3,438 \\
Urea Nitrogen [Re.] & $\geq$20 mg/dL  & 41.6\% &  240,892  \\
Albumin [Me.] & $\leq$3.5 g/dL  & 22.5\% &  57,040 \\
Hemoglobin [He.] & $\leq$13.7 g/dL  & 72.39\% & 236,802 \\
Cholesterol HDL [Me.] & $\leq$41 mg/dL  &  23.31\% & 9,023 \\
RDW-SD [He.] & $\geq$46.3 fL  & 49.16\% & 73,001  \\
Absolute Basophil Count [Im.] & $\leq$0.01 K/uL  & 7.81\% & 51,315 \\
Creatinine [Re./Ca.] & $\leq$0.5 mg/dL  & 1.03\% &  241,968  \\
Red Blood Cells [He.] & $\leq$4.6 m/uL  & 73.4\% & 234079   \\
RDW [He.] & $\geq$15.5\%  & 22.55\% & 234,022  \\
Hematocrit [He.] & $\leq$40\%  & 67.77\% & 238,567   \\
INR PT [Co.] & $\geq$1.1 ratio  & 46.15\% & 137,973 \\
Bilirubin, Total [Me.] & $\geq$1.5 mg/dL  & 10.36\% & 75,016   \\
PT [Co.] & $\geq$12.5 sec  & 54.11\% & 137,917 \\
Lymphocytes [Im.] & $\leq$18\%  & 42.77\%  & 171,441 \\
MCHC [He.] & $\leq$31 g/dL  & 8.61\%  & 234,103  \\ \hline\hline
\end{tabular}
\label{table:est_prev_sup}
\end{table*}

\begin{table*}[ht]
\caption{Label prevalence (Prev.) and sample count (Counts) for the labels investigated in the main manuscript across the tasks laboratory values abnormality prediction. Next to each value name, we highlight the physiological category it represents: Cardiac (Ca.), Renal (Re.), Hematological (He.), Metabolic (Me.), Immunological (Im.), and Coagulation (Co.). The values in bold represent the highest performance per abnormality. For PT at the 30~minute horizon no score was estimated as it did not satisfy the inclusion criterion of at least 10 abnormal cases in both test and validation set.}
\centering
\begin{tabular}{lcccc} \hline\hline
\textbf{Value} & \textbf{Threshold}  & \textbf{30m. Prev./Counts} & \textbf{60m. Prev./Counts} & \textbf{120m. Prev./Counts} \\ \hline
NTproBNP [Ca.] & $\geq$353 pg/mL & 78.93\% / 11672 & 77.47\% / 17398 & 75.95\% / 23174 \\
Hemoglobin [He.] & $\geq$17.5 g/dL  & 0.38\% / 105483 & 0.34\% / 149731 & 0.31\% / 197263 \\
Hematocrit [He.] & $\geq$51\% & 0.62\% / 104851 & 0.56\% / 149341 & 0.52\% / 197743 \\
Urea Nitrogen [Re.] & $\leq$6 mg/dL & 0.81\% / 104076 & 0.8\% / 152037 & 0.88\% / 203206 \\
Creatinine [Re./Ca.] & $\geq$1.2 mg/dL & 28.53\% / 104595 & 28.46\% / 152946 & 28.43\% / 204543 \\
Urea Nitrogen [Re.] & $\geq$20 mg/dL  & 41.83\% / 104076 & 41.71\% / 152037 & 41.47\% / 203206 \\
Red Blood Cells [He.] & $\geq$6.1 m/uL  & 0.34\% / 103427 & 0.3\% / 146919 & 0.29\% / 193343 \\
Hemoglobin [He.] & $\leq$13.7 g/dL & 70.59\% / 105483 & 71.71\% / 149731 & 72.90\% / 197263 \\
Albumin [Me.] & $\leq$3.5 g/dL & 22.96\% / 23560 & 22.47\% / 34015 & 22.87\% / 46419 \\
RDW-SD [He.] & $\geq$46.3 fL & 49.56\% / 27755 & 48.85\% / 44519 & 48.54\% / 65870 \\
Absolute Monocyte Count [Im.] & $\leq$0.2 K/uL & 3.49\% / 20660 & 3.24\% / 33509 & 3.20\% / 49559 \\
RDW [He.] & $\geq$15.5\%   & 21.86\% / 103402 & 22.02\% / 146880 & 22.49\% / 193284 \\
Absolute Lymphocyte Count [Im.] & $\geq$ 3.7 K/uL & 3.46\% / 20660 & 3.06\% / 33509 & 2.87\% / 49552 \\
Red Blood Cells [He.] & $\leq$4.6 m/uL & 71.71\% / 103427 & 72.48\% / 146919 & 73.37\% / 193343 \\
C-Reactive Protein [Im.] & $\geq$ 5 mg/L & 63.09\% / 1303 & 58.63\% / 2195 & 57.95\% / 3241 \\
Hematocrit [He.] & $\leq$40\% & 66.08\% / 104851 & 66.98\% / 149341 & 68.03\% / 197743 \\
Absolute Basophil Count [Im.] & $\leq$0.01 K/uL & 5.63\% / 20660 & 5.27\% / 33509 & 5.11\% / 49556 \\
INR PT [Co.] & $\geq$1.1 ratio  & 44.25\% / 60491 & 45.51\% / 83393 & 46.88\% / 109475 \\
Lymphocytes [Im.] & $\geq$42\%  & 5.71\% / 81887 & 5.58\% / 114732 & 5.65\% / 147267 \\
PT [Co.] & $\geq$12.5 sec & 53.6\% / 60409 & 54.09\% / 83310 & 54.65\% / 109393 \\
Lymphocytes [Im.] & $\leq$18\% & 42.54\% / 81887 & 41.66\% / 114732 & 41.11\% / 147267 \\
Anion Gap [Me.] & $\geq$20 mEq/L  & 8.09\% / 100624 & 7.64\% / 146795 & 7.55\% / 196396 \\
MCHC [He.] & $\leq$31 g/dL & 11.3\% / 103446 & 10.69\% / 146935 & 10.16\% / 193356 \\
Bilirubin Total [Me.] & $\geq$1.5 mg/dL & 10.06\% / 30934 & 9.76\% / 44888 & 9.83\% / 60625 \\ \hline\hline
\end{tabular}
\label{table:mon_prev_sup}
\end{table*}

\end{document}